\documentclass[journal]{IEEEtran}
\usepackage{nomencl}
\makenomenclature
\usepackage{etoolbox}
\usepackage{xstring}
\usepackage{amsmath,amsfonts}
\usepackage{algorithmicx}
\usepackage[ruled,vlined,linesnumbered]{algorithm2e}
\SetNlSty{text}{}{} 
\RestyleAlgo{ruled}  
\SetAlgorithmName{Algorithm}{Algorithm}{List of Algorithms}
\usepackage{array}
\usepackage{eqparbox} 
\usepackage[caption=false,font=normalsize,labelfont=sf,textfont=sf]{subfig}
\usepackage{textcomp}
\usepackage{stfloats}
\usepackage{url}
\usepackage{verbatim}
\usepackage{graphicx}
\usepackage{cite}
\usepackage{multirow}
\usepackage{algpseudocode}
\usepackage{hyperref}
\usepackage{amsthm}
\theoremstyle{plain}

\hypersetup{colorlinks=true,
            linkcolor=blue,
            anchorcolor=blue,
            citecolor=blue}

\renewcommand{\nomgroup}[1]{%
  \item[\bfseries
  \ifstrequal{#1}{A}{Abbreviations}{%
  \ifstrequal{#1}{G}{Greek Symbols}{%
  \ifstrequal{#1}{R}{Roman Symbols}{}}}%
]}

\begin{document}

\title{An End-to-End Approach for Microgrid Probabilistic Forecasting and Robust Operation via Decision-focused Learning}

\author{Tingwei Cao, and Yan Xu,~\IEEEmembership{Senior Member,~IEEE}
\thanks{}
\thanks{}}

\markboth{}%
{}

\IEEEpubid{}

\maketitle

\begin{abstract}
High penetration of renewable energy sources (RES) introduces significant uncertainty and intermittency into microgrid operations, posing challenges to economic and reliable scheduling.  To address this, this paper proposes an end-to-end decision-focused framework that jointly optimizes probabilistic forecasting and robust operation for microgrids. A multilayer encoder–decoder (MED) probabilistic forecasting model is integrated with a two-stage robust optimization (TSRO) model involving direct load control (DLC) through a differentiable decision pathway, enabling gradient-based feedback from operational outcomes to improve forecasting performance.  Unlike conventional sequential approaches, the proposed method aligns forecasting accuracy with operational objectives by directly minimizing decision regret via a surrogate smart predict-then-optimize (SPO) loss function. This integration ensures that probabilistic forecasts are optimized for downstream decisions, enhancing both economic efficiency and robustness. Case studies on modified IEEE 33-bus and 69-bus systems demonstrate that the proposed framework achieves superior forecasting accuracy and operational performance, reducing total and net operation costs by up to 18\% compared with conventional forecasting and optimization combinations. The results verify the effectiveness and scalability of the end-to-end decision-focused approach for resilient and cost-efficient microgrid management under uncertainty.
\end{abstract}



\section{Introduction}
\IEEEPARstart{T}{he}
microgrids have emerged as a pivotal component of modern power systems, facilitating the integration of distributed RES such as PV and WT, along with energy storage systems (ESS) and demand-side management mechanisms. By operating either in grid-connected or islanded modes, microgrids enhance energy flexibility, reliability, and resilience against external disturbances \cite{zhang2024two,fei2025weather}. However, the high penetration of RES and user-side load demand stochasticity introduces inherent intermittency and uncertainty \cite{che2015optimal,tiwari2018design}. These uncertainties significantly challenge the optimal scheduling and operation of microgrids, leading to potential inefficiencies and stability risks.

Over the past decade, a substantial body of research has focused on optimizing microgrid operation through coordinated control of ESS, distributed generators, and demand-side resources \cite{imani2018demand,aaslid2022stochastic}. Deterministic optimization methods initially provided foundational scheduling strategies under nominal scenarios, while robust and stochastic optimization frameworks were later developed to explicitly address uncertainty in RES generation and load demand. Despite these advances, traditional optimization frameworks often rely on pre-defined uncertainty models or exogenous forecasts, which may not adequately capture the complex data-driven patterns and decision interactions inherent in microgrid operation \cite{dawoud2018hybrid,hu2023economic}.

To effectively characterize and mitigate the impact of uncertainty in microgrid operation, various data-driven probabilistic forecasting methods have been proposed \cite{xie2017variable,ye2019data,kalhori2022data}. Early studies typically assumed known uncertainty intervals or distributional bounds for renewable generation and load demand, enabling tractable robust or chance constrained formulations \cite{zhang2016robust}. Subsequent developments employed statistical time series models to capture temporal dependencies and quantify prediction intervals \cite{zhang2017robust}. More recently, advances in machine learning have driven the adoption of neural network based probabilistic forecasting approaches, including recurrent architectures, attention based models, and generative methods, which achieve superior accuracy in capturing nonlinear temporal spatial correlations \cite{wang2018combining,hu2022self,cao2024feature}. While these methods have substantially improved forecasting accuracy, they remain largely decoupled from downstream decision making. Forecast models are typically trained to minimize statistical metrics without considering their operational consequences, leading to suboptimal or even inconsistent performance when integrated into optimization tasks. Consequently, the gap between probabilistic forecasting and decision-focused optimization limits the practical effectiveness of current microgrid management frameworks under uncertainty \cite{sang2022safety,wang2023acceleration}.
\IEEEpubidadjcol

To bridge this gap between forecasting accuracy and operational performance, recent studies have explored the integration of forecasting and optimization into unified decision making frameworks \cite{el2019generalization,elmachtoub2020decision,elmachtoub2022smart}. However, most existing approaches still follow a sequential pipeline, where forecasting is performed independently before being fed into an optimization model. Moreover, conventional robust or stochastic optimization methods often assume fixed uncertainty distributions. Motivated by these challenges \cite{li2021optimal,vu2022optimal}, this paper proposes an end-to-end decision focused framework that jointly optimizes probabilistic forecasting and robust operation for microgrids. Specifically, a multilayer probabilistic forecasting module is coupled with a TSRO model through a differentiable decision pathway, enabling gradient-based feedback from operational performance to forecasting model training. This unified structure aligns statistical learning with decision objectives, ensuring that the generated forecasts are directly optimized for downstream operational robustness and economic efficiency.

The main contributions of this study are summarized as follows:
\begin{enumerate}
    \item A novel forecasting and optimization paradigm is first proposed for microgrid operation under uncertainty, fully considering renewable intermittency, load fluctuation, and coordinated control of ESS and load demand side.
    \item Different from traditional forecasting then optimization schemes, this paper develops a decision-focused forecasting model, which can learn from both data and operational feedback, ensuring that forecasting accuracy and decision performance are optimized simultaneously.
    \item Compared with conventional forecasting and optimization combinations, the proposed end-to-end approach achieves significantly improved operational economy and robustness.
\end{enumerate}

The remainder of this paper is organized as follows: Section II introduces the overall framework and system formulation; Section III details the probabilistic forecasting model; Section IV presents the robust optimization model and solution algorithm; Section V provides case studies; and Section VI concludes the paper.

\section{End-to-end Approach for Microgrid Probabilistic Forecasting and Robust Operation}
A microgrid is an integrated local energy system that comprises several key physical components, each playing a critical role in enabling efficient and resilient power supply. As depicted in the figure, the microgrid includes RES generation, such as photovoltaic panels and wind turbines, which produce clean electricity but introduce intermittency and uncertainty due to their dependence on weather conditions. The ESS acts as a buffer, storing excess energy during periods of high RES output and discharging it to meet demand when generation is low, thereby enhancing flexibility and reliability. The electrical network interconnects these components, facilitating power flow and distribution, while the load represents end-user electricity consumption that exhibits variability and unpredictability. Additionally, the microgrid is coupled with the main grid, allowing for energy exchange to balance supply-demand gaps and improve overall stability. However, the inherent uncertainties from RES generation and load demand pose significant challenges to microgrid operation.

To achieve optimal operation performance of microgrids, this paper seeks to coordinate ESS and DLC to cooperatively handle RES and load demand uncertainties.

\begin{figure}[ht]
    \centering
    \includegraphics[width=8.5cm]{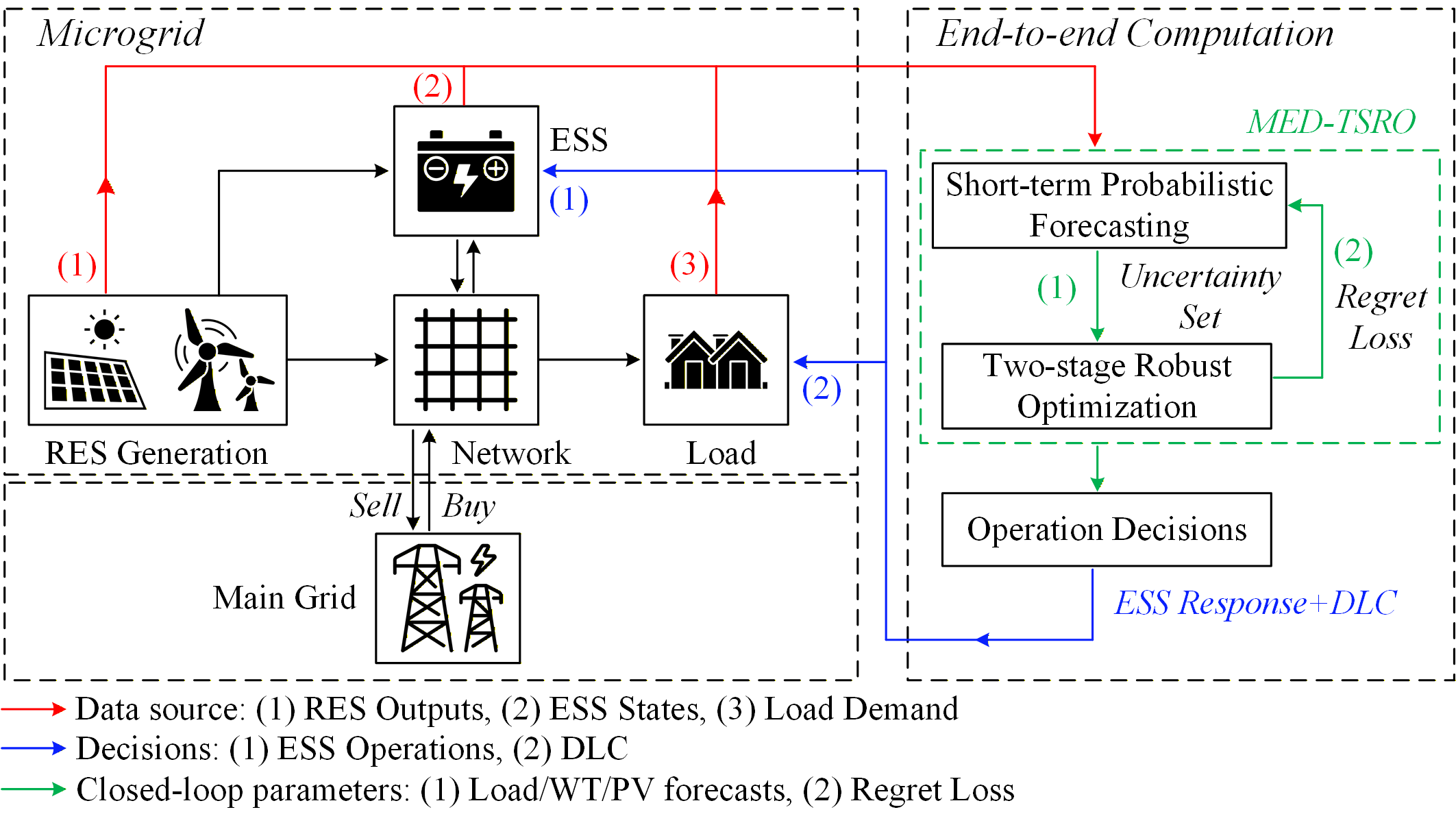}
    \caption{Structure of the end-to-end probabilistic forecasting and robust optimization approach for microgrid}
    \label{fig:mos}
\end{figure}

The proposed end-to-end framework in this paper integrates probabilistic forecasting and robust optimization into a unified pipeline, addressing the inherent uncertainties in renewable energy generation and load demand. As illustrated in Fig. \ref{fig:mos}, the architecture consists of two core components: a decision-focused probabilistic forecasting module and a TSRO module, which are synergistically coupled through a differentiable decision-making pathway. This holistic approach enables simultaneous optimization of forecasting accuracy and operational performance, overcoming the limitations of traditional decoupled methods where forecasting errors may propagate adversely into operational decisions.

Fig. \ref{fig:mos} illustrates the comprehensive workflow of the proposed end-to-end framework, beginning with historical data ingestion and probabilistic forecasting through the MED model, proceeding to uncertainty set construction and robust optimization via the TSRO model, and culminating in operational decision execution. The bidirectional arrow between forecasting and optimization modules represents the gradient-based coupling mechanism that enables joint training of both components. This integrated architecture effectively bridges the gap between statistical forecasting and decision-making, ensuring that microgrid operators can achieve economically efficient and robust performance under uncertainty.

\section{Decision-focused Short-term Probabilistic Forecasting}
\subsection{Proposed MED Methodology}
This study proposes a novel yet computationally efficient architecture based on Multilayer Perceptrons (MLPs) for short-term time series forecasting tasks.

\begin{figure}[ht]
    \centering
    \includegraphics[width=8.5cm]{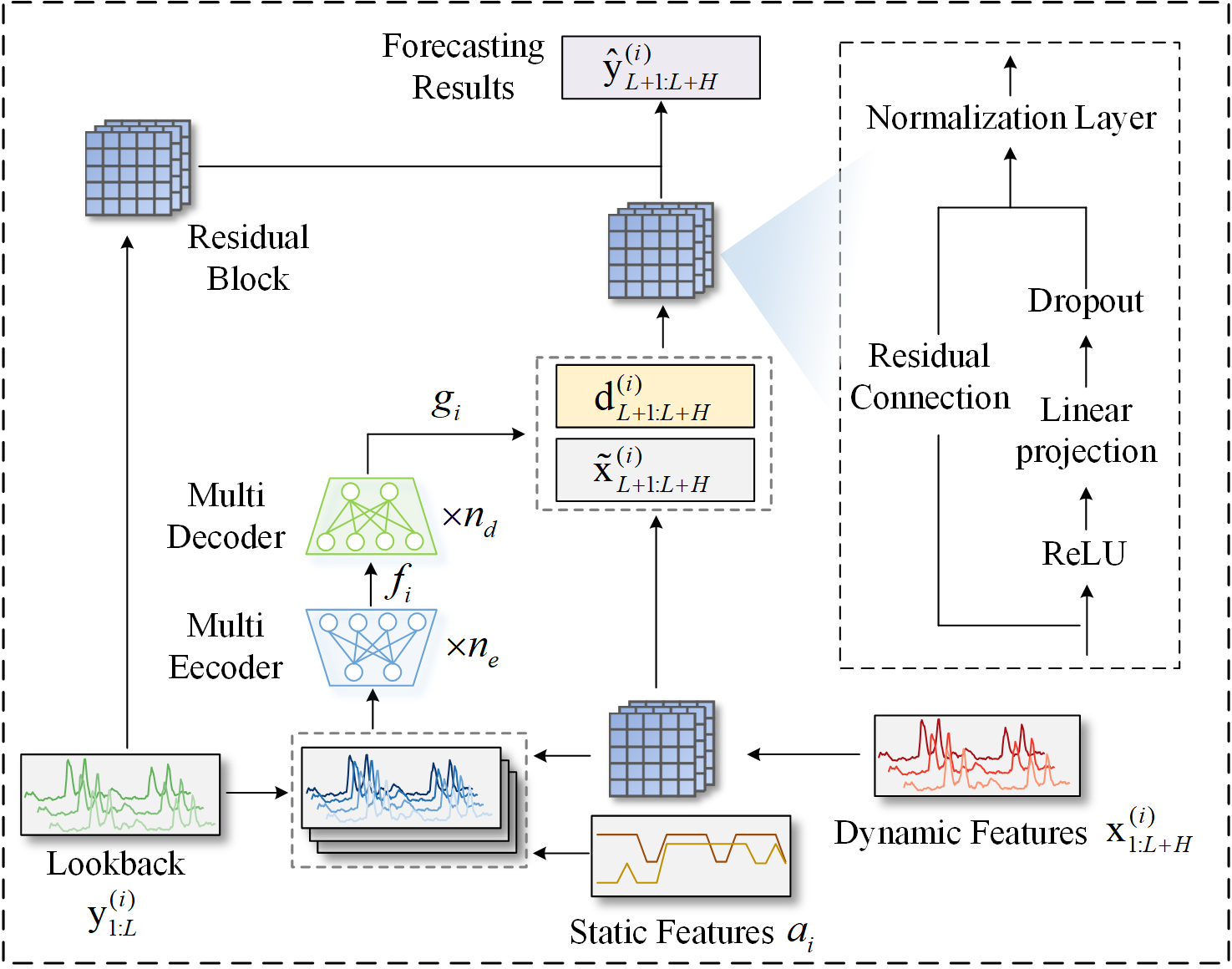}
    \caption{The overall framework of the MED Model}
    \label{fig:TiDE-DA}
\end{figure}

\subsubsection{MED Model Framework}
The proposed MED framework adopts a hierarchical architecture, as illustrated in Fig. \ref{fig:TiDE-DA}.

The raw inputs are first projected into a unified latent space via residual blocks. Each temporal feature vector undergoes nonlinear transformation through the residual operation defined in \eqref{relu}-\eqref{dropout}.

The transformed features are flattened and concatenated with historical demand and static attributes to form a fused context tensor as \eqref{c0}. A stack of $\{ {f_i}\} _{i = 1}^{{n_e}}$ encoder layers   then progressively compresses ${c_0}$ into a high-level representation $e$ as \eqref{ci}. The final output combines decoded temporal patterns with residual historical signals as \eqref{rhs}.

\subsubsection{Residual Block}
Given an input vector  ${\mathbf{x}} \in {\mathbb{R}^{{d_{{\text{in}}}}}}$, the block employs a sequential transformation: First, the input is projected to a hidden dimension  ${d_{{\text{hid}}}}$ via a learnable linear layer. This is followed by an element-wise ReLU activation as defined in \eqref{relu}. Subsequently, a second linear layer compresses the hidden representation back to the original input dimension through the transformation specified in \eqref{2}.
\begin{equation}
    {\mathbf{h}} = {\text{ReLU}}({{\mathbf{W}}_1}{\mathbf{x}} + {{\mathbf{b}}_1}),{{\mathbf{W}}_1} \in {\mathbb{R}^{{d_{{\text{hid}}}} \times {d_{{\text{in}}}}}},{{\mathbf{b}}_1} \in {\mathbb{R}^{{d_{{\text{hid}}}}}}
    \label{relu}
\end{equation}
\begin{equation}
    {\mathbf{z}} = {{\mathbf{W}}_2}{\mathbf{h}} + {{\mathbf{b}}_2},{{\mathbf{W}}_2} \in {\mathbb{R}^{{d_{{\text{in}}}} \times {d_{{\text{hid}}}}}},{{\mathbf{b}}_2} \in {\mathbb{R}^{{d_{{\text{in}}}}}}
    \label{2}
\end{equation}

To enhance gradient flow and stabilize training, a residual connection adds the original input to the transformed output. Mathematically, the processed output ${\mathbf{O}}$ can be expressed as:
\begin{equation}
    {\mathbf{O}} = {\text{Dropout}}\left( {{\text{LayerNorm}}\left( {{\mathbf{X}} + \mathcal{F}({\mathbf{X}})} \right)} \right)
    \label{dropout}
\end{equation}

\subsubsection{Multiple Encoder and Decoder}
The proposed architecture employs parallel encoder-decoder chains to process heterogeneous temporal features. Let ${{\mathbf{y}}_{1:L}} \in {\mathbb{R}^{{d_y}}}$ denote historical power demand and ${{\mathbf{X}}_L} \in {\mathbb{R}^{{d_x}}}$ represent external features at time $t$. The encoder first transforms raw inputs through residual blocks:
\begin{equation}
    {{\mathbf{\tilde x}}_t} = {\mathcal{R}_\theta }({{\mathbf{x}}_t}),\;\;\;{\kern 1pt} \forall t \in 1, \ldots ,L + H
    \label{xt}
\end{equation}

${\mathcal{R}_\theta }$ denotes the residual transformation in \eqref{relu}-\eqref{dropout}. The flattened features $\widetilde {\mathbf{X}} \in {\mathbb{R}^{(L + H) \times {d_h}}}$ are concatenated with historical load demand and auxiliary data ${\mathbf{a}}$:
\begin{equation}
    {{\mathbf{c}}_0} = [{{\mathbf{y}}_{1:L}}|\widetilde {\mathbf{X}}|{\mathbf{a}}] \in {\mathbb{R}^{{d_c}}}
    \label{c0}
\end{equation}

Multiple encoding layers $\{ {f_i}\} _{i = 1}^{{n_e}}$ then progressively compress the fused representation:
\begin{equation}
    {{\mathbf{c}}_i} = {f_i}({{\mathbf{c}}_{i - 1}}),\;\;\;{\kern 1pt} {f_i}( \cdot ) = \sigma ({{\mathbf{W}}_i}( \cdot ) + {{\mathbf{b}}_i})
    \label{ci}
\end{equation}

$\sigma $ denotes the Swish activation function. The final projection combines decoded features with residual historical patterns:
\begin{equation}
    {{\mathbf{\hat y}}_{L + 1:L + H}} = \sum\limits_{t = 1}^H {{\mathcal{T}_\phi }} ([{{\mathbf{d}}_t}|{{\mathbf{\tilde x}}_{L + t}}]) + {{\mathbf{W}}_{{\text{res}}}}{{\mathbf{y}}_{1:L}}
    \label{rhs}
\end{equation}

${\mathcal{T}_\phi }$ denotes the temporal decoder with attention mechanism for temporal alignment, and ${{\mathbf{W}}_{{\text{res}}}}$ enables direct propagation of historical trends.

\subsection{Loss Function Reformulation}

\subsubsection{Statistical Error Metric}
The Continuous Ranked Probability Score (CRPS) is a scoring rule used to evaluate the accuracy of continuous probability predictions. A lower CRPS indicates a better estimation, and the formula is as follows:
\begin{equation}
    CRPS = \frac{1}{T}\sum\limits_{t = 1}^T {\int_{ - \infty }^{ + \infty } {{{[F({{\hat y}_t}) - H({{\hat y}_t} - {y_t})]}^2}d{{\hat y}_t}} }
\end{equation}

${F({{\hat y}_t})}$ is the cumulative distribution function of the predictive distribution, and ${H({{\hat y}_t} - {y_t})}$ is the Heaviside step function of the observation $y$.

When the predictive distribution is represented by the sample $\{ {y_1}, \cdots ,{y_n}\} $, CRPS can be approximated as \eqref{CRPS}.
\begin{equation}
    \label{CRPS}
    CRPS \approx \frac{1}{T}\sum\limits_{i = 1}^T {\left| {{{\hat y}_i} - \bar y} \right|}  - \frac{1}{{2{T^2}}}\sum\limits_{i = 1}^T {\sum\limits_{j = 1}^T {\left| {{{\hat y}_i} - {{\hat y}_j}} \right|} }
\end{equation}

Note that the sample-based CRPS in \eqref{CRPS} involves absolute-value terms; in our implementation we use automatic differentiation with subgradients at the non-differentiable points, which does not affect gradient-based training in practice.

\subsubsection{Surrogate Decision-focused Loss and Differentiability}
Traditional probabilistic metrics such as CRPS only measure statistical accuracy, while microgrid operation performance is determined by the downstream optimization decisions.
Let $D$ denote the realized uncertain quantity, and $\hat D$ denote the forecast output. We write the downstream operation problem in a compact parametric form as
\begin{equation}
J(D)=\min_{z\in \mathcal{Z}(D)} f(z;D),
\quad z=(x,y)
\label{10}
\end{equation}
where $x$ and $y$ represent the first and second stage operation variables, respectively, and $\mathcal{Z}(D)$ denotes the feasible set induced by network, ESS and DLC constraints.

When the operation model contains binary variables or non-smooth constraints, the optimal value function $J(\cdot)$ is in general non-differentiable and may even be discontinuous. Hence, directly using $\nabla J(\hat D)$ for end-to-end training is not mathematically justified.

To obtain a well-defined gradient pathway, we introduce a continuous convex surrogate of the operation layer by relaxing binary variables to $[0,1]$ and applying standard convex relaxations to nonconvex constraints, yielding a convex feasible set $\bar{\mathcal{Z}}(D)$, and adding a Tikhonov regularization term with $\rho>0$ to ensure strong convexity:
\begin{equation}
    \bar z_\rho(D)=\arg\min_{z\in \bar{\mathcal{Z}}(D)} f(z;D) + \frac{\rho}{2}\|z\|_2^2
    \label{zd}
\end{equation}
\begin{equation}
    \bar J_\rho(D)=f(\bar z_\rho(D);D)+\frac{\rho}{2}\|\bar z_\rho(D)\|_2^2
\label{jd}
\end{equation}

This surrogate is always solvable under mild feasibility and boundedness conditions, and admits a unique optimizer due to $\rho$-strong convexity.

We define a regret-type decision-focused loss by comparing the operational performance under the forecast-induced decision
against the oracle decision under the realized $D$:
\begin{equation}
L_{\mathrm{SPO}}(D,\hat D)= f\!\left(\bar z_\rho(\hat D);\,D\right)-f\!\left(\bar z_\rho(D);\,D\right)
\label{12}
\end{equation}

To improve numerical stability and keep the loss nonnegative and smooth, we further use the softplus operator:
\begin{equation}
    \mathrm{Regret}_{\mathrm{SPO}}(D,\hat D)=\log\!\left(1+\exp\!\left(L_{\mathrm{SPO}}(D,\hat D)\right)\right)
    \label{regret_spo}
\end{equation}

During training, gradients are taken through the surrogate optimizer $\bar z_\rho(\hat D)$ in \eqref{zd}. Let the Karush-Kuhn-Tucker (KKT) residual of \eqref{zd} and \eqref{jd} be denoted by $F(\bar z_\rho,\lambda,\nu;\hat D)=0$ (stationarity, primal feasibility and complementarity).
If the Jacobian $\partial F/\partial(\bar z_\rho,\lambda,\nu)$ is nonsingular, then $\bar z_\rho(\hat D)$ is differentiable in a neighborhood and
\begin{equation}
    \frac{\partial(\bar z_\rho,\lambda,\nu)}{\partial \hat D}
    =-\left(\frac{\partial F}{\partial(\bar z_\rho,\lambda,\nu)}\right)^{-1}
    \left(\frac{\partial F}{\partial \hat D}\right)
    \label{1111}
\end{equation}
which yields $\nabla_{\hat D}\mathrm{Regret}_{\mathrm{SPO}}(D,\hat D)$ by the chain rule.

The gradients are backpropagated using \eqref{1111} only through the continuous surrogate optimizer in \eqref{zd}–\eqref{jd}. The TSRO solved by C\&CG is used for forward decision evaluation and benchmarking, which is not differentiated.

\section{Robust Operation Model}
This section introduces the robust operation model used to optimize microgrid performance under uncertainty. The model leverages TSRO to account for uncertainty in the system's components. Additionally, this section presents the operation constraints, including energy storage and demand-side management, and discusses the solution approach through the Column-and-Constraint Generation (C\&CG) algorithm for improved computational efficiency.

\subsection{Objective Function}
The proposed optimization model aims to maximize net economic benefit, formulated as total revenue from electricity sales to the demand customers minus the OM costs of ESS, renewable energy integration, and transaction with main grid and loads. Based on the conclusion of \cite{zhang2016robust}, this article sets the DLC at one quarter of an hour. It aims to modify the load demands when the RES outputs have significant deviations from the hour-ahead expected values.

According to \cite{poonpun2008analysis,zhang2016robust}, all the costs of the energy storage system, including the investment cost, replacement cost, and charging/discharging cost, can be converted into a maintenance cost that is used for the purpose of simplifying the analysis, expressed as $C_{\text{ESS,OM}}$ (\$/MWh).

The ESS maintenance cost is equivalently modeled as a unified operational cost coefficient ${C_{{\text{ESS,OM}}}}$, converted from charging/discharging efficiency and degradation costs.

Binary variables $\alpha_{\text{ch},m,j},\alpha_{\text{dis},m,j} \in {0,1}$ indicate the charging/discharging states of ESS units at node $m$, constrained by rated power limits. We use $P_{{\text{dis,}m}}^{\max }/P_{{\text{ch,}m}}^{\max }$ to represent the rated operating power of this node, and ${L_{{\text{dis,}}m,j}}/{L_{{\text{ch,}}m,j}}$ to represent the charging/discharging power rate of ESS unit at node $m$ on level $j$.

On the other hand, in the DLC model, the total load ${P_D}$ is decomposed into controllable and uncontrollable parts with ratio ${K_{D,{\text{con}}}}$.

Based on the model established above, the operation objective function is formulated as follows,
\begin{align}
    \label{o1} \text{min } \;C_{\mathrm{ESS}} + &C_{\mathrm{WT}} + C_{\mathrm{PV}} + C_{\mathrm{Grid}} - C_{\mathrm{rev}}\\
    \text{s.t. } \;C_{\text{ESS}} = &C_{\text{ESS,ch}}\sum_{m \in N_{\text{ESS}}} P_{\text{ch},m}^{\max} \sum_{j \in J_{\text{ch}}} \alpha_{\text{ch},m,j} L_{\text{ch},m,j} \nonumber \\
    + &C_{\text{ESS,dis}}\sum_{m \in N_{\text{ESS}}} P_{\text{dis},m}^{\max} \sum_{j \in J_{\text{dis}}} \alpha_{\text{dis},m,j} L_{\text{dis},m,j} \\
    C_{\text{WT}} =& C_{\text{WT,OM}}\sum_{n \in N_{\text{WT}}} \sum_{t \in T} \frac{P_{\text{WT},n,t}}{N_T} \\
    C_{\text{PV}} =& C_{\text{PV,OM}}\sum_{n \in N_{\text{PV}}} \sum_{t \in T} \frac{P_{\text{PV},n,t}}{N_T} \\
    C_{\text{Grid}} =& C_{\text{buy}}\sum_{t \in T} \frac{P_{def,t}}{N_T} - C_{\text{sell}}\sum_{t \in T} \frac{P_{sur,t}}{N_T} \\
    C_{\text{rev}} =& C_{\text{rev,con}} + C_{\text{rev,unc}} \nonumber \\
    =& p_{D,\text{con}}\sum_{q \in HQ} (K_{D,\text{con}} - K_{\text{DLC},q}) \sum_{i \in N_D} \frac{P_{D,i}}{4} \nonumber \\
    +& p_{D,\text{unc}}(1 - K_{D,\text{con}})\sum_{i \in N_D} P_{D,i}
\end{align}

The decision-focused term is used only for training the forecasting model. Specifically, the training loss is
\begin{equation}
\min_{\theta}\;\mathbb{E}\Big[\mathrm{CRPS}(\theta)\;+\;\lambda\cdot \mathrm{Regret}_{\mathrm{SPO}}(D,\hat D_\theta)\Big]
\label{oa}
\end{equation}
where $\hat D_\theta$ is the forecast output of MED with parameters $\theta$, and $\lambda$ is selected by grid search.

\subsection{Operation Constraints}
Power flow and voltage relations follow the linearized DistFlow model \cite{zhang2017robust}, ensuring nodal power balance and voltage drop limits.
\begin{align}
    P_{ij} &= g_{ij}(u_i - u_j) - b_{ij}\theta_{ij} + \ell_{ij}r_{ij} \label{eq:pflow} \\
    Q_{ij} &= -b_{ij}(u_i - u_j) - g_{ij}\theta_{ij} + \ell_{ij}x_{ij} \label{eq:qflow} \\
    \ell_{ij} &= \frac{P_{ij}^2 + Q_{ij}^2}{u_i} \quad 
    \label{eq:2}
\end{align}
\begin{equation}
    \left\lVert 
    \begin{bmatrix} 
        2P_{ij} \\ 
        2Q_{ij} \\ 
        \ell_{ij} - u_i 
    \end{bmatrix} 
    \right\rVert_2 \leq \ell_{ij} + u_i, \quad \forall ij \in L
    \label{eq:3}
\end{equation}
\begin{equation}
    u_i - u_j = 2(r_{ij}P_{ij} + x_{ij}Q_{ij}) - (r_{ij}^2 + x_{ij}^2)\ell_{ij}
    \label{eq:voltagedrop}
\end{equation}

$\ell_{ij}$ is the line current square, which needs to be transformed into a convex constraint by relaxation. An auxiliary variable $\ell_{ij}$ is introduced to convert the non-linear term into a second-order cone constraint as \eqref{eq:3}. And \eqref{eq:voltagedrop} is the voltage drop equation.

\begin{align}
P_{b+1,t} &= P_{b,t} - P_{0,b+1,t} - P_{\mathrm{ch},m}^{\max} \sum_{j \in J_{\mathrm{ch}}} \alpha_{\mathrm{ch},m,j} L_{\mathrm{ch},m,j} \nonumber \\
&+ P_{\mathrm{dis},m}^{\max} \sum_{j \in J_{\mathrm{dis}}} \alpha_{\mathrm{dis},m,j} L_{\mathrm{dis},m,j} - P_{\mathrm{DLC},i,q} \nonumber \\
&+ P_{\mathrm{WT},n,t} + P_{\mathrm{PV},n,t}, \quad b \in Br(i), \forall i,t,q \label{eq:22}
\end{align}
\begin{equation}
Q_{b+1,t} = Q_{b,t} - Q_{0,b+1,t} - Q_{\mathrm{DLC},i,q}, b \in Br(i), \forall i,t,q \label{eq:23}
\end{equation}
\begin{equation}
V_{i+1,t} = V_{i,t} - \frac{R_{b} P_{b,t} + X_{b} Q_{b,t}}{V_{0}}, b \in Br(i,i+1), \forall i,t \label{eq:24}
\end{equation}

\begin{equation}
1-V^{\max } \leq V_{i, t} \leq 1+V^{\max }, \forall i, t
\label{27}
\end{equation}
\begin{equation}
P_{1, t}=P_{\mathrm{def}, t}-P_{\mathrm{sur}, t}, P_{\mathrm{def}, t} \geq 0, P_{\mathrm{sur}, t} \geq 0, \forall t
\label{28}
\end{equation}

Operational constraints enforce binary charging/discharging exclusivity, ESS energy balance, renewable generation bounds, and DLC adjustment limits.

\subsection{Robust Operation Model}
In order to enable the solution to cope with the uncertainties of renewable energy output power and load demand in the future, we have developed a TSRO model to coordinate the operation of ESS and DLC at different time scales. This model ensures operational feasibility under any realization of the uncertainty within a predefined set, thus enhancing the system's resilience.

\subsubsection{TSRO Model}
The proposed microgrid coordination model can be converted to a TSRO model, which adopts a two-stage ``{\it min–max–min}'' structure, where first-stage variables $x$ represent pre-dispatch decisions, second-stage variables $y$ capture recourse actions, and $u$ denotes uncertain renewable and load realizations:
\begin{align}
    \label{ojf1}\min \limits_x \quad&{c^T}x + \mathop {\max }\limits_u \mathop {\min }\limits_y {d^T}y + {e^T}u\\
    \label{ojf2}\text{s.t.}\quad&Ax \geq b\\
    \label{ojf3}&Fx + Gy + Hu \leq v\\
    \label{ojf4}&I(u)x + Jy + Ku = w\\
    \label{ojf5}&u \in \mathcal{U}
\end{align}

Assume that (i) the uncertainty set $\mathcal{U}$ is nonempty, compact and polyhedral;
(ii) the first-stage feasible set $\mathcal{X}$ is nonempty and compact; and
(iii) the second-stage recourse set is nonempty and bounded for every $(x,u)\in\mathcal{X}\times\mathcal{U}$.

Under the above assumption, the TSRO problem admits an optimal solution and has a finite optimal value. Moreover, the differentiable surrogate problem in \eqref{zd}-\eqref{jd} admits a unique optimizer for any input due to $\rho>0$.

Therefore, we classify the three variables in the two stages as $x$, $y$ and $u$. First, $x$ represents a set of decision variables in the first stage operation, including ESS daily charging and discharging binary variables ${\alpha _{\text{ch}}}$, ${\alpha _{\text{dis}}}$, and DLC contracts for the entire scheduling period ${K_{D,\text{con}}}$. Secondly, $y$ stands for a group of adjustable variables in the second stage including the quarter-hour ahead DLC controllable variables. ${K_{\text{DLC}}}$ for each quarter hour can be optimized in the second stage after the realization of the uncertain variables, and they are referred as the “wait-and-see” decisions. Thirdly, $u$ stands for the uncertainty variables which are the outputs of the wind turbines and the PVs, and the uncertain load demand $P_D$.

To characterize the uncertainties of renewable generation and load demand, we define the following polyhedral uncertainty sets:
\begin{equation}
    \mathcal{U}_{\mathrm{WT}} = \left\{ P_{\mathrm{WT},n,t} \left| 
\begin{array}{l}
P_{\mathrm{WT},n,t} = \hat{P}_{\mathrm{WT},n,t} + \Delta P_{\mathrm{WT},n,t}, \\
\sum_{t \in T} \left| \frac{\Delta P_{\mathrm{WT},n,t}}{\Delta P_{\mathrm{WT},n}^{\max}} \right| \leq \Gamma_{\mathrm{WT}}, \\
\left| \Delta P_{\mathrm{WT},n,t} \right| \leq \Delta P_{\mathrm{WT},n}^{\max}, \quad \forall n,t
\end{array}
\right. \right\}
\label{uwt}
\end{equation}

\begin{equation}
    \mathcal{U}_{\mathrm{PV}}=\left\{P_{\mathrm{PV},n,t}\left|\begin{array}{l}
P_{\mathrm{PV},n,t}=\hat{P}_{\mathrm{PV},n,t}+\Delta P_{\mathrm{PV},n,t},\\
\sum_{t\in T}\left|\frac{\Delta P_{\mathrm{PV},n,t}}{\Delta P_{\mathrm{PV},n}^{\max }}\right|\leq\Gamma_{\mathrm{PV}},\\
\left|\Delta P_{\mathrm{PV},n,t}\right|\leq\Delta P_{\mathrm{PV},n}^{\max },\quad\forall n,t
\end{array}\right.\right\}
\label{upv}
\end{equation}
\begin{equation}
    \mathcal{U}_{D}=\left\{P_{D, i} \left| 
\begin{array}{l}
P_{D, i}=\hat{P}_{D, i}+\Delta P_{D, i}, \\
\sum_{i \in N_{D}}\left|\frac{\Delta P_{D, i}}{\Delta P_{D, i}^{\max }}\right| \leq \Gamma_{D}, \\
\left|\Delta P_{D, i}\right| \leq \Delta P_{D, i}^{\max }, \quad \forall i
\end{array}
\right. \right\}
\label{ud}
\end{equation}

The parameters $\Gamma_{\mathrm{WT}}$, $\Gamma_{\mathrm{PV}}$ and $\Gamma_{\mathrm{D}}$ serve as uncertainty budgets, controlling the conservativeness of the optimization by limiting the cumulative normalized deviations across time or nodes. For simplicity, we use the uncertain budget parameter $\Gamma = \frac{\Gamma_{\mathrm{WT}}}{\left| T \right|} = \frac{\Gamma_{\mathrm{PV}}}{\left| T \right|} = \frac{\Gamma_{\mathrm{D}}}{\left| N_D \right|}$ to represent these three parameters. This structure ensures computational tractability while enabling the model to immunize against worst-case realizations within the prescribed bounds, thereby enhancing operational resilience without excessive cost.

The unified objective \eqref{o1} is reformulated in the TSRO form as \eqref{o2}.
\begin{align}
\label{o2}
  \mathop {\min }\limits_{{\alpha _{\text{ch}}},{\alpha _{\text{dis}}}} {C_{\text{ESS}}} &+ \mathop {\max }\limits_{{P_{\text{WT}}},{P_{\text{PV}}}} \mathop {\min }\limits_{K_{\text{DL}C},V,P,Q} {C_{\text{WT}}} + {C_{\text{PV}}} \nonumber\\
   &+ {C_{\text{Grid}}} - {C_{rev}}
\end{align}

In the first stage, the variables ${\alpha _{ch}}$ and ${\alpha _{dis}}$ for energy storage are optimized to minimize the cost $C_{\text{ESS}}$. The max operation then considers the worst-case realizations of $P_{\text{WT}}$ and $P_{\text{PV}}$ from these uncertainty sets, capturing adversarial scenarios. Finally, the inner min operation adjusts operational variables $K_{\text{DLC}}$, $V$, $P$ and $Q$ to minimize the total cost, including $C_{\text{WT}}+C_{\text{PV}}$ and grid exchange costs $C_{\text{grid}}-C_{\text{rev}}$.

\subsubsection{Column-and-Constraint Generation (C\&CG) Algorithm}
As demonstrated in \cite{zeng2013solving}, the C\&CG algorithm exhibits significantly faster convergence compared to traditional Benders decomposition \cite{bertsimas2012adaptive}. Therefore, this paper intends to employ the C\&CG method to solve this problem.

However, in the proposed end-to-end forecasting and optimization framework, the integration of the SPO loss term which enables gradient-based coupling between forecasting and decision making introduces a need for modifications to the standard C\&CG method.

Traditional C\&CG operates under a decomposition strategy that iterates between a master problem (MP) and a subproblem (SP). The MP is a relaxation of the full TSRO, while the SP identifies the worst-case scenario from the uncertainty set for a given first-stage solution. However, when the objective includes a regret term $\lambda \cdot \text{Regret}_{\text{SPO}}$, which is itself a function of the forecast model parameters and the decision variables, the conventional algorithm does not consider the gradient propagation from the operational cost back to the forecast model. This limits the ability to fully implement an end-to-end learning pipeline.

To address these issues, we solve TSRO using the standard C\&CG procedure, and update the forecasting model using the surrogate SPO gradient.

The algorithm begins by initializing the iteration counter $k=0$ and establishing the initial uncertainty set $\mathcal{U}^{(0)}$ using nominal forecast values ${\hat P_{\text{WT}}}$, ${\hat P_{\text{PV}}}$ and ${\hat P_{D}}$. This initial set represents the baseline uncertainty realization before considering worst-case scenarios. The convergence threshold $\epsilon$ is set to $10^{-4}$ to ensure sufficient solution precision while maintaining computational efficiency.

The MP constitutes a relaxed version of the original TSRO formulation, incorporating all identified worst-case scenarios from previous iterations, which provides the optimal first-stage decision $x^*$ and establishes a lower bound for the overall objective value.
\begin{align}
    \min_{x, y_1, \ldots, y_k, \eta}& c^T x + \eta\\
    \text{s.t.\;\;\;\;} &\eta  \geq d^T y_i + e^T u_i \nonumber \\
    &A x  \geq b \\
    &F x + G y_i + H u_i  \leq \nu \\
    &I(u_i) x + J y_i + K u_i  = w, \forall i=1, \ldots, k \\
    &x  \in \mathcal{X}, y_i \in \mathcal{Y}
\end{align}

The SP aims to identify the worst-case uncertainty realization for the given first-stage solution $x^*$:
\begin{align}
    \Phi\left(x^{*}\right)&=\max _{u\in\mathcal{U}}\min _{y} \;d^{T}y+e^{T}u \nonumber \\
    \text{s.t.\;\;\;} &Fx^{*}+Gy+Hu\leq v\\
    &I(u)x^{*}+Jy+Ku=w\\
    &y\in\mathcal{Y}
\end{align}

To slove the SP, this paper utilize the KKT optimality conditions and Big-M method to convert the bi-level problem into a MILP model. The reformulation becomes:
\begin{align}
    \max_{u,y,\mu,\nu}\;\; &d^{T}y + e^{T}u\\
    \text{s.t.}\;\;& u \in \mathcal{U} \\
    & Fx^{*} + Gy + Hu \leq v \\
    & I(u)x^{*} + Jy + Ku = w \\
    & d + G^{T}\mu + J^{T}\nu = 0 \\
    & \mu \geq 0 \\
    & \label{cc}\mu_{i}(v_{i} - F_{i}x^{*} - G_{i}y - H_{i}u) = 0, \quad \forall i
\end{align}

The Big-M method provides an effective approach to linearize the nonlinear complementarity constraints arising from the KKT conditions-based reformulation.

For each complementarity constraint in \eqref{cc}, the nonlinear constraint can then be replaced by the following set of linear constraints:
\begin{align}
    &\mu_{i} \leq M_{i} \delta_{i} \\
    &v_{i} - F_{i} x^{*} - G_{i} y - H_{i} u \leq M_{i} (1 - \delta_{i}) \\
    &\mu_{i} \geq 0 \\
    &\delta_{i} \in \{0, 1\}
\end{align}
where $\delta_{i} \in \{0, 1\}$ is a a binary variable and $M_i > 0$ is a sufficiently large constant.

\begin{algorithm}[ht]
\label{alg:1}
\caption{End-to-End Training Architecture}
\KwIn{Historical data $\mathbf{Y}=\{P_{PV}, P_{WT}, P_D\}$, external features $\mathbf{X}$, learning rate $\gamma$, convergence threshold $\varepsilon$}
\KwOut{Optimized MED parameters $\theta^*$, optimal scheduling decisions $x^*, y^*$}

\textbf{1. Initialization:} \\
MED network parameters $\theta^{(0)}$; \\
Uncertainty sets $\mathcal{U}^{(0)}_{WT}, \mathcal{U}^{(0)}_{PV}, \mathcal{U}^{(0)}_D$; \\
Iteration counter $k = 0$;

\vspace{0.5em}
\textbf{2. Forward Propagation:} \\

Step 1: Generate probabilistic forecasts using MED \\
\hspace{1em} $(\hat{\mathbf{Y}}^{(k)}, \Sigma^{(k)}) = \text{MED}(\mathbf{Y}, \mathbf{X}; \theta^{(k)})$;

Step 2: Construct uncertainty sets\\
\hspace{1em} $(\mathcal{U}_{WT}, \mathcal{U}_{PV}, \mathcal{U}_D) = \text{UncertaintySets}(\hat{\mathbf{Y}}^{(k)}, \Sigma^{(k)})$;

Step 3: Solve TSRO via C\&CG (forward pass only): \\
\hspace{1em} \textit{(a) MP:} \\
\hspace{2em} $\min_{x,\eta} \; c^\top x + \eta$ \\
\hspace{1em} \textit{(b) SP:} \\
\hspace{2em} $\max_{u} \min_{y} \left(d^\top y + e^\top u\right)$ \\
\hspace{1em} Iterate MP and SP until $|UB-LB| < \varepsilon$; \\[0.2em]
\hspace{1em} Obtain optimal cost $J^*(\hat{\mathbf{Y}}^{(k)})$ and decisions;

Step 4: Compute decision regret $\text{Regret}_{\text{SPO}}$: \\
\hspace{1em} $L_{\mathrm{SPO}}^{(k)} = f\!\left(\bar z_\rho(\hat Y^{(k)});\,Y\right)-f\!\left(\bar z_\rho(Y);\,Y\right)$,\\
\hspace{1em}$\mathrm{Regret}_{\mathrm{SPO}}^{(k)}=\log\!\left(1+\exp(L_{\mathrm{SPO}}^{(k)})\right)$;

\vspace{0.5em}
\textbf{3. Backward Propagation:} \\

Step 5: Obtain $\nabla_{\hat Y}\mathrm{Regret}_{\mathrm{SPO}}^{(k)}$ (surrogate layer only):\\
\hspace{1em}$\nabla_{\theta} L_{\mathrm{SPO}}^{(k)} = \left(\nabla_{\hat Y}\mathrm{Regret}_{\mathrm{SPO}}^{(k)}\right)^{\top}\nabla_{\theta}\hat Y^{(k)}$;

Step 6: Update MED parameters: \\
\hspace{1em} $\theta^{(k+1)}=\theta^{(k)}-\gamma(\nabla_{\theta}\mathrm{CRPS}+\lambda\,\nabla_{\theta}L_{\mathrm{SPO}}^{(k)})$;

\vspace{0.5em}
\textbf{4. Iteration and Convergence:} \\
If $|\mathcal{L}_{SPO}^{(k+1)} - \mathcal{L}_{SPO}^{(k)}| < \varepsilon$, stop; otherwise set $k = k + 1$ and return to Step 1.
\end{algorithm}

\subsubsection{End-to-end Training Architecture}
The end-to-end training architecture proposed in this study integrates the MED and the TSRO in a unified pipeline, enabling joint optimization of forecasting accuracy and operational performance. The framework establishes a differentiable pathway between the two models, allowing gradients to flow through the surrogate decision layer during training, and ensuring that both forecasting and optimization tasks are aligned with decision-making objectives.

Algorithm \ref{alg:1} illustrates the complete training procedure, which involves four primary steps: initialization, forward propagation, backward propagation, and iteration. In the first step, the MED network parameters are initialized, and the uncertainty sets $\mathcal{U}_{WT}$, $\mathcal{U}_{PV}$, and $\mathcal{U}_D$ are defined based on nominal forecast values. 

The forward propagation step begins by generating probabilistic forecasts using the MED model. The forecasting output includes both predicted values and associated uncertainty intervals, represented as $\hat{\mathbf{Y}}^{(k)}$ and $\Sigma^{(k)}$ respectively. These forecasts are then used to construct updated uncertainty sets in step 3, which define the ranges of possible future renewable generation and load demand realizations. The step 4 solves the TSRO problem using the C\&CG algorithm, with the objective of minimizing the operational cost while considering the uncertainty sets.

In the backward propagation phase, the decision regret is computed by comparing the actual operational cost with the expected optimal cost $J^*(D)$. The gradient of the regret is then propagated back through the system, updating the parameters of the MED forecasting model based on the operational feedback. And then the algorithm continues iterating until the convergence condition is met, which is when the change in the SPO loss between iterations falls below a specified threshold $\epsilon$.

The end-to-end architecture enables simultaneous learning of both the forecasting model and the robust optimization model by backpropagating decision gradients from the optimization results to the forecasting model.

\section{Case Study}
\subsection{Dataset Description and Environment Setting}
This section introduces the datasets utilized for training and evaluating the proposed MED forecasting model and robust optimization framework.

The dataset includes multivariate time-series data collected from urban microgrid networks with a temporal resolution of 15-min intervals \cite{medtsro}. Key features are categorized as follows:

Daily load demand and PV/WT power, serving as the target
variable for forecasting. Season labels, month labels, weekend labels, daily maximum and minimum temperatures, weather condition index, sourced from meteorological databases. Real time gasoline prices, obtained from regional energy authorities. Solar irradiance index, measured from on-site pyranometers.

\begin{figure}[ht]
    \centering
    \includegraphics[width=8.5cm]{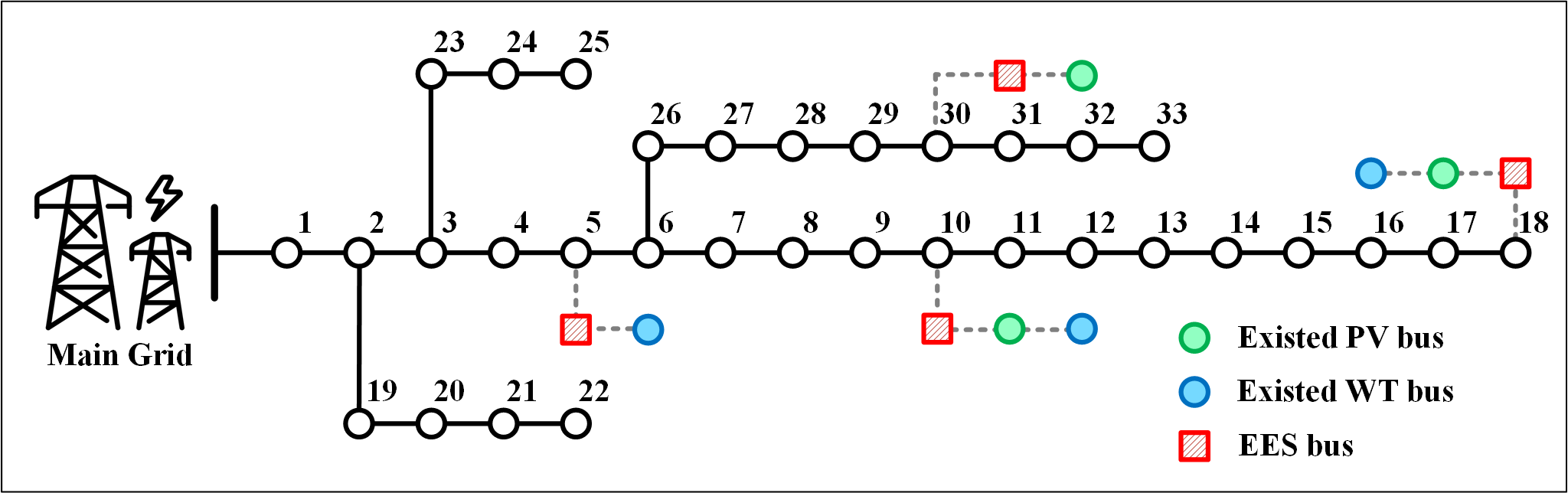}
    \caption{Modified IEEE 33-bus PDN}
    \label{fig:IEEE 33-BUS}
\end{figure}

\begin{figure}[ht]
    \centering
    \includegraphics[width=8.5cm]{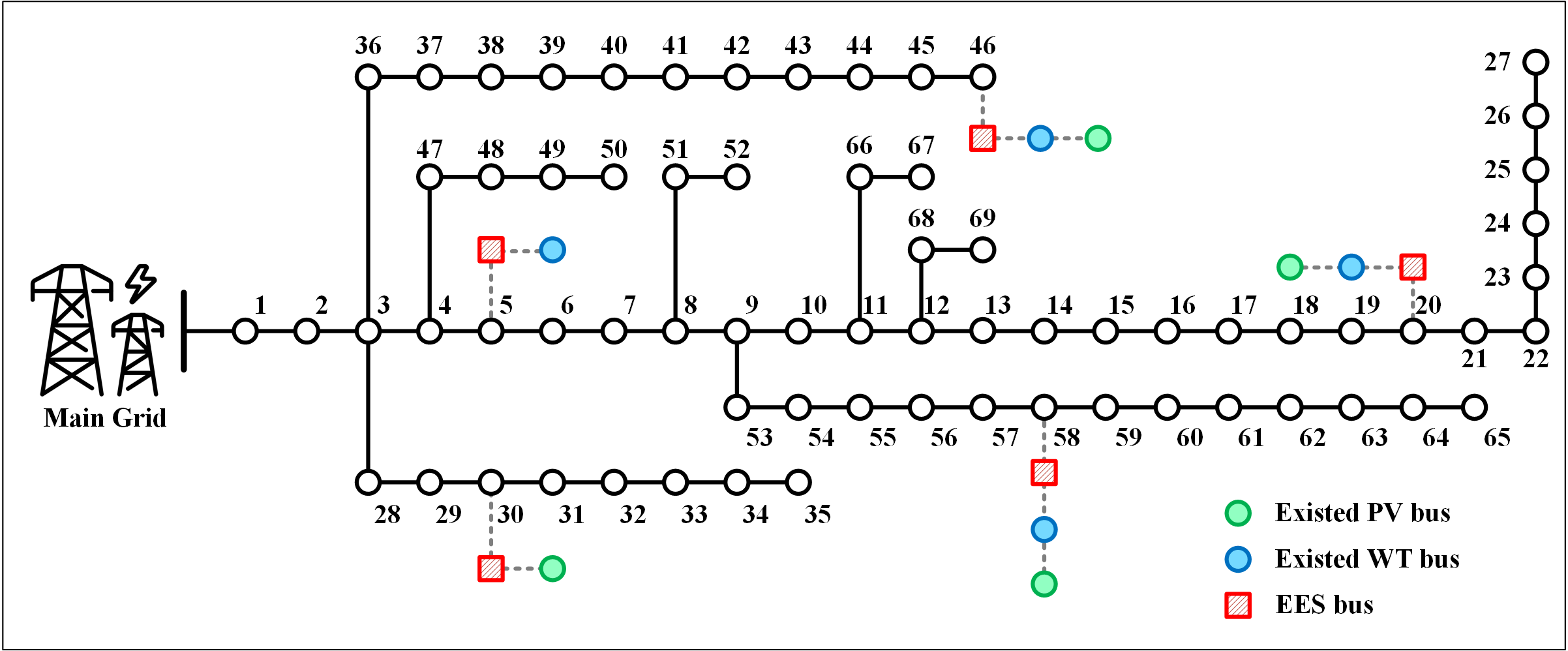}
    \caption{Modified IEEE 69-bus PDN}
    \label{fig:IEEE 69-BUS}
\end{figure}

To support the optimization model’s physical constraints, the dataset includes topology parameters for the modified IEEE 33-bus and 69-bus power distribution networks (PDNs). These include: Bus voltage limits, line impedance values (resistance and reactance), and thermal capacity thresholds. Existing renewable energy integrations (e.g., PV and WT capacities at designated buses), visualized as Fig. \ref{fig:IEEE 33-BUS} and Fig. \ref{fig:IEEE 69-BUS}. The rated power of ESS and RES nodes, and the rated stored energy parameters of ESS nodes are shown in Table \ref{tab:para_essres}. Table \ref{tab:para_2} lists the parameters for microgrid forecasting and operation tests.

The case studies are conducted on a 64-bit Windows PC with AMD Core R7-9700X CPU@3.80GHz of 16 cores and 32 GB RAM, and Geforce Nvidia RTX-4080 GPU. All codes and data processing scripts are implemented in Python 3.8.18 with reproducible random seeds to ensure result consistency.

\begin{table}[ht]
\centering
\renewcommand{\arraystretch}{1.3}
\caption{Parameters of ESS and RES Nodes}
\resizebox{\linewidth}{!}{
\begin{tabular}{ccccc}
\hline
Network                 & \begin{tabular}[c]{@{}c@{}}Node\\ Type\end{tabular} & \begin{tabular}[c]{@{}c@{}}Node\\ No.\end{tabular} & \begin{tabular}[c]{@{}c@{}}Rated Power\\ (MWh)\end{tabular} & \begin{tabular}[c]{@{}c@{}}Rated Stored Energy\\ (MWh)\end{tabular} \\ \hline
\multirow{3}{*}{33-Bus} & ESS                                                 & 5,10,18,30                                         & 1.5                                                         & 2                                                                   \\
                        & WT                                                  & 5,10,18                                            & 1                                                           & --                                                                   \\
                        & PV                                                  & 10,18,30                                           & 1                                                           & --                                                                  \\
\multirow{3}{*}{69-Bus} & ESS                                                 & 5,20,30,46,58                                      & 1.5                                                         & 2                                                                   \\
                        & WT                                                  & 5,20,46,58                                         & 1                                                           & --                                                                   \\
                        & PV                                                  & 20,30,46,58                                        & 1                                                           & --                                                                   \\ \hline 
\end{tabular}}
\label{tab:para_essres}
\end{table}

\begin{table}[ht]
\centering
\renewcommand{\arraystretch}{1.3}
\caption{Parameters for Microgrid Forecasting and Operation Tests}
\resizebox{\linewidth}{!}{
\begin{tabular}{cccccc}
\hline
Parameter & \begin{tabular}[c]{@{}c@{}}Value\\ (\$/kWh)\end{tabular} & Parameter & \begin{tabular}[c]{@{}c@{}}Value\\ (\$/kWh)\end{tabular} & Parameter & Value        \\ \hline
$C_{\text{cell}}$                             & 0.015                                                    & $C_{\text{ESS,dis}}$                             & 0.05                                                     & $\eta_{\text{ch}}$                             & 0.9          \\
$C_{\text{buy}}$                             & 0.07                                                     & $C_{\text{ESS,ch}}$                             & -0.02                                                    & $\eta_{\text{dis}}$                             & 1.2          \\
$p_{D,\text{con}}$                             & 0.06                                                     & $C_{\text{WT,OM}}$                             & 0.01                                                     & $\beta$                             & 0.9          \\
$p_{D,\text{unc}}$                             & 0.1                                                      & $C_{\text{PV,OM}}$                             & 0.01                                                     & $\lambda$                             & {[}0.1,10{]} \\ \hline
\end{tabular}}
\label{tab:para_2}
\end{table}

\subsection{Probabilistic Forecasting Performance}
\begin{figure}[ht]
    \centering
    \includegraphics[width=8.5cm]{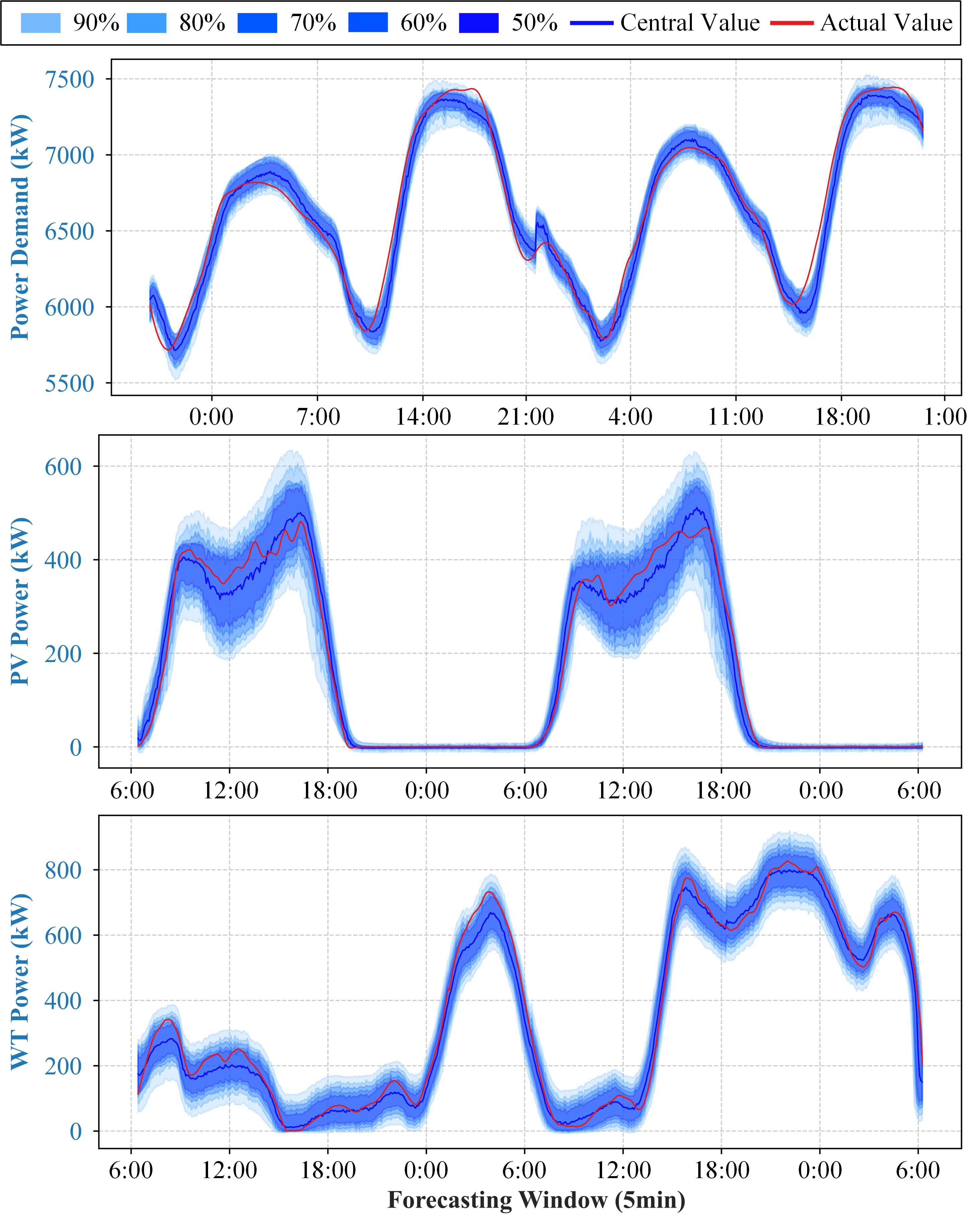}
    \caption{Examples of the result of probabilistic forecasting achieved by MED with SPO Loss}
    \label{fig:forecast}
\end{figure}

This subsection evaluates the probabilistic forecasting performance of the proposed MED model, which integrates the SPO loss function for decision-focused learning. The forecasting results are visualized in Fig. \ref{fig:forecast}, providing a clear depiction of the model's ability to capture uncertainties in load demand and renewable energy generation.

As shown in Fig. \ref{fig:forecast}, the MED model generates probabilistic forecasts for three variables, including PV power, WT power, and load demand, over a 5-minute resolution timeline. The plots exhibit the predicted probability distributions through shaded confidence intervals and point forecasts (median values), compared against actual observed data points. Notably, the MED model demonstrates high accuracy in tracking temporal patterns, with narrow prediction intervals indicating reduced epistemic uncertainty, particularly during peak demand periods and intermittent renewable generation phases. This performance is attributed to the hierarchical architecture of MED, which leverages residual blocks and encoder-decoder chains to fuse historical data with external features. Moreover, the incorporation of the SPO loss function ensures that forecasting errors are minimized in a decision-aware manner, leading to improved alignment with downstream operation.

\begin{table}[ht]
\centering
\renewcommand{\arraystretch}{1.3}
\caption{Comparison of Forecasting Performance with Baseline Models}
\resizebox{\linewidth}{!}{
\begin{tabular}{llcccc}
\hline
Model       & Dataset & \multicolumn{1}{l}{R-square} & \multicolumn{1}{l}{RMSE} & \multicolumn{1}{l}{CRPS} & \multicolumn{1}{l}{Pinball Loss} \\ \hline
MED         & Demand  & 0.9651                       & 94.76                    & 0.72                     & 0.073                            \\
            & PV      & 0.9843                       & 23.52                    & 0.64                     & 0.068                            \\
            & WT      & 0.9852                       & 33.52                    & 0.67                     & 0.069                            \\
DeepAR      & Demand  & 0.9432                       & 147.62                   & 1.04                     & 0.098                            \\
            & PV      & 0.9671                       & 56.87                    & 0.96                     & 0.095                            \\
            & WT      & 0.9535                       & 67.93                    & 0.93                     & 0.094                            \\
Transformer & Demand  & 0.9521                       & 153.28                   & 1.12                     & 0.112                            \\
            & PV      & 0.9594                       & 51.18                    & 0.92                     & 0.094                            \\
            & WT      & 0.9647                       & 64.39                    & 0.91                     & 0.092                            \\
Bi-LSTM     & Demand  & 0.9363                       & 189.33                   & 1.37                     & 0.131                            \\
            & PV      & 0.9527                       & 62.43                    & 1.13                     & 0.112                            \\
            & WT      & 0.9498                       & 83.52                    & 1.24                     & 0.125                            \\ \hline
\end{tabular}}
\label{tab_compar1}
\end{table}

Table \ref{tab_compar1} shows a comprehensive evaluation of the proposed MED model against three baseline models (DeepAR, Transformer, and Bi-LSTM \cite{salinas2020deepar,wang2022transformer,cui2022district}) across three distinct datasets: load demand, PV power, and WT power. The performance is assessed using four key metrics: R-square, RMSE, CRPS, and Pinball Loss.

The table clearly demonstrates that the MED model consistently achieves superior performance compared to all baseline models. For example, on the load demand dataset, MED attains an R-square of 0.9651, significantly higher than the best baseline (Transformer at 0.9521), along with lower error metrics such as an RMSE of 94.76 versus Transformer's 153.28. This trend holds across all datasets and metrics; for instance, on PV power, MED's R-square is 0.9843, outperforming DeepAR's 0.9671, and on WT power, MED shows a Pinball Loss of 0.069, which is lower than any baseline. Overall, the MED model exhibits enhanced predictive accuracy and reliability, highlighting its effectiveness in forecasting tasks for renewable energy and demand scenarios.

\subsection{Robust Operation Outcomes}
\subsubsection{Economic Efficiency Analysis}
The robust operation results presented in this subsection evaluate the performance of the proposed end-to-end framework under varying uncertainty levels. By leveraging the probabilistic forecasts generated by the MED model, the TSRO model solved via the C\&CG algorithm determines the optimal scheduling of ESS, DLC, and power exchange with the main grid.
\begin{figure}[ht]
    \centering
    \includegraphics[width=8.5cm]{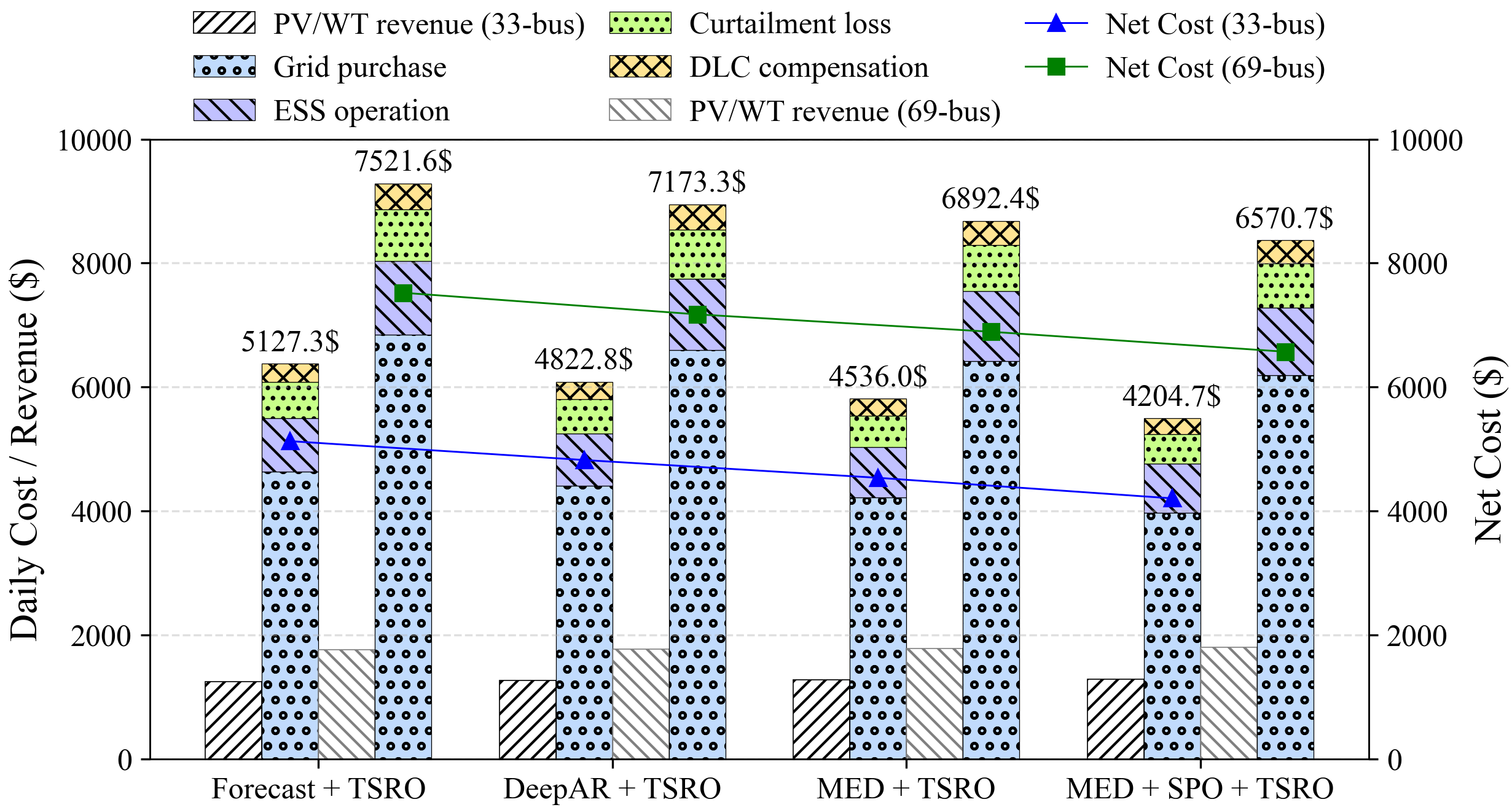}
    \caption{Composition of daily operation cost and revenue for the microgrids under different forecasting–optimization schemes}
    \label{fig:Cost_Composition_33_69bus}
\end{figure}

We conducted a comparison of several forecasting and optimization combination frameworks. These frameworks include the combination of deterministic forecasting and DeepAR with TSRO, as well as the two-stage forecasting then optimization framework (MED + TSRO), and the end-to-end framework (MED + SPO + TSRO).

Fig. \ref{fig:Cost_Composition_33_69bus} illustrates the composition of daily operation cost and revenue for both the 33-bus and 69-bus microgrids with $\Gamma = 0.1$. Each stacked column represents the cost contributions from grid electricity purchase, ESS operation, renewable curtailment loss, and DLC compensation, while the hatched segments denote the distributed generation revenue from PV and wind units. Across both systems, the proposed end-to-end framework reduced by approximately 18\% on average total and net operating costs. This improvement is particularly evident in the 69-bus system, demonstrating scalability under larger network sizes. The reduction arises mainly from more balanced grid-exchange scheduling and efficient ESS utilization, enabled by the decision-focused joint training that aligns probabilistic forecasts with the downstream robust-optimization objective.
\begin{table*}[ht]
\caption{Operation cost breakdown and net cost for microgrids under four forecasting–optimization schemes}
\label{tab:operation_costs}
\centering
\renewcommand{\arraystretch}{1.3}
\begin{tabular}{l c c c c c c c r}
\hline
Method & System & Grid (\$) & ESS (\$) & Curtailment (\$) & DLC (\$) & PV Rev (\$) & Total (\$) & Net Cost (\$) / Reduction \\
\hline
Forecast + TSRO         & 33-bus & 4,632.4 & 861.7 & 583.6 & 298.5 & 1,248.9 & 6,376.2 & 5,127.3 / --- \\
DeepAR + TSRO          & 33-bus & 4,411.6 & 833.8 & 557.9 & 283.2 & 1,263.7 & 6,086.5 & 4,822.8 / 5.9\% \\
MED + TSRO      & 33-bus & 4,223.8 & 812.6 & 503.4 & 271.4 & 1,275.2 & 5,811.2 & 4,536.0 / 11.5\% \\
MED + SPO + TSRO       & 33-bus & 3,975.3 & 791.4 & 468.9 & 258.6 & 1,289.5 & 5,494.2 & 4,204.7 / 18.0\% \\
\hline
Forecast + TSRO         & 69-bus & 6,848.5 & 1,186.2 & 827.8 & 421.6 & 1,762.5 & 9,284.1 & 7,521.6 / --- \\
DeepAR + TSRO          & 69-bus & 6,595.2 & 1,155.5 & 792.4 & 404.3 & 1,774.1 & 8,947.4 & 7,173.3 / 4.6\% \\
MED + TSRO      & 69-bus & 6,420.7 & 1,126.8 & 739.6 & 391.2 & 1,785.9 & 8,678.3 & 6,892.4 / 8.4\% \\
MED + SPO + TSRO       & 69-bus & 6,188.9 & 1,097.2 & 706.1 & 376.9 & 1,798.4 & 8,369.1 & 6,570.7 / 12.6\% \\
\hline
\end{tabular}
\end{table*}

Table \ref{tab:operation_costs} provides a detailed quantitative breakdown of the daily operation results corresponding to Fig. 6. In the 33-bus system, the proposed end-to-end method lowers the net cost from \$5127 to \$4205, with a reduction of approximately 18\%, while in the 69-bus microgrid, the improvement reaches about 12.6\%. The MED-based probabilistic forecasts already reduce unnecessary conservatism in the robust model, and the inclusion of the SPO decision-focused learning further enhances economic performance by explicitly minimizing operational regret during training.

\begin{figure}[ht]
    \centering
    \includegraphics[width=8.5cm]{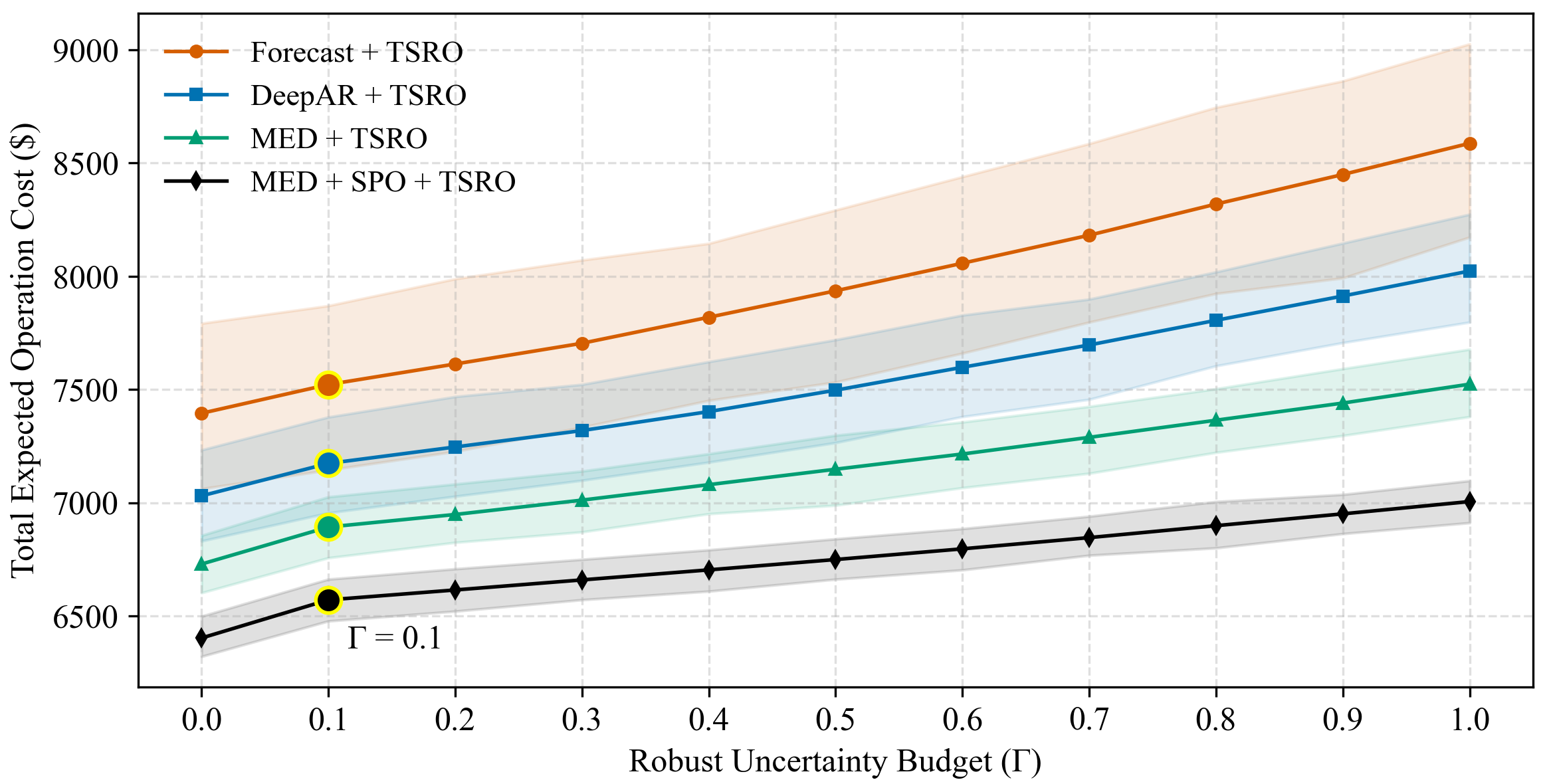}
    \caption{Total operation cost under different robust uncertainty budget in 69-bus microgrid}
    \label{fig:Cost_vs_Gamma_MonteCarloCI}
\end{figure}

\subsubsection{Robustness Evaluation}
Fig. \ref{fig:Cost_vs_Gamma_MonteCarloCI} presents the variation of total expected operation cost of 69-bus microgrid with the robust uncertainty budget $\Gamma$ for all forecasting–optimization schemes, with shaded bands indicating the 95\% confidence intervals derived from Monte Carlo simulations of forecast uncertainty realizations. The uncertainty budget jointly represents the variability in load demand, photovoltaic, and wind power forecasts. The highlighted point at $\Gamma=0.1$, which represents a typical uncertainty level. As $\Gamma$ increases, all methods exhibit higher operating costs, but the degree of increase and the width of confidence intervals differ markedly. The baseline Forecast + TSRO shows the steepest growth and the largest variability, indicating higher sensitivity to uncertainty, whereas the proposed MED + SPO + TSRO maintains both the lowest mean cost, and the 95\% CI width reduced by 42\% compared with baseline.

\subsubsection{Dispatch Interpretability}
\begin{figure}[ht]
    \centering
    \includegraphics[width=8.5cm]{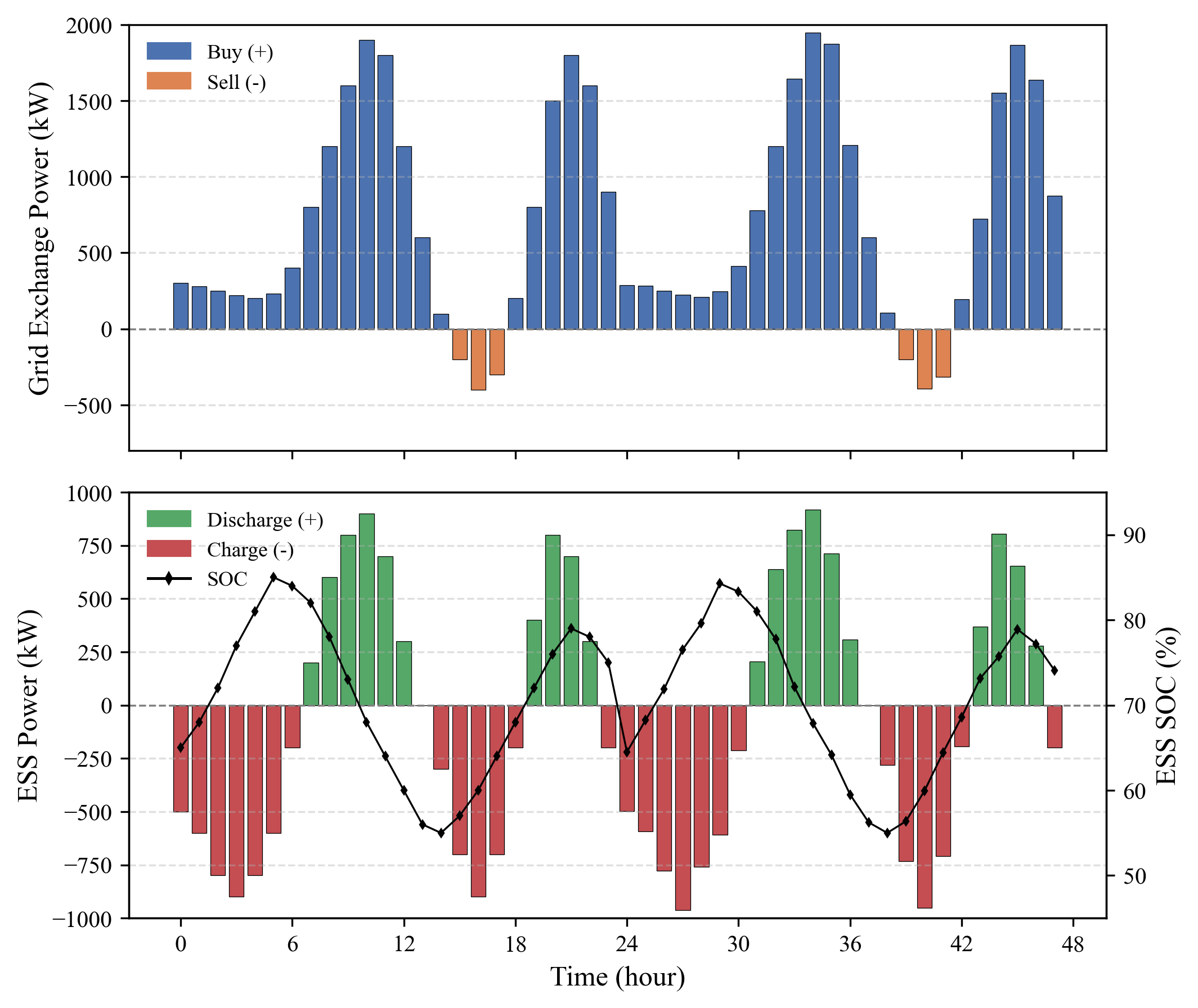}
    \caption{Time-series dispatch profiles of ESS and grid exchange}
    \label{fig:MED_SPO_TSRO_48h}
\end{figure}

Fig. \ref{fig:MED_SPO_TSRO_48h} illustrates the 48-hour operation of the microgrid under the proposed MED + SPO + TSRO framework, providing detailed profiles of grid-exchange power, ESS dispatch, and the corresponding SOC.

In the upper subplot, the grid-exchange power is represented by bars, where positive values denote electricity purchased from the main grid and negative values indicate power exported during renewable surpluses. Two daily purchase peaks around 09:00 and 19:00 align with local demand peaks, while negative bars near noon reflect photovoltaic generation exceeding local consumption.

The lower subplot presents the ESS operation, where the bars describe charging and discharging power, and the line on the right axis represents the SOC trajectory. The ESS charges during nighttime and solar-rich hours and discharges during load peaks, maintaining SOC levels between 55\% and 85\% throughout the 48-hour horizon.

These coordinated power-exchange and storage behaviors confirm that the proposed end-to-end approach achieves both economic efficiency and operational robustness. It also implicitly learns cost-aware temporal dependencies between renewable curtailment and storage flexibility, an ability absent in decoupled forecasting then optimization.

\section{Conclusion}
This paper presented an end-to-end framework that integrates probabilistic forecasting and robust operation for microgrids. By coupling the MED forecasting model with the TSRO via a differentiable SPO loss, the proposed method bridges the gap between predictive learning and operational decision-making. The MED model effectively captures multi-source temporal dependencies and delivers calibrated uncertainty estimates, while the TSRO model, solved by a modified column-and-constraint generation algorithm, ensures resilient scheduling under renewable and demand uncertainties.

Extensive case studies on 33-bus and 69-bus microgrids verify that the proposed end-to-end framework significantly enhances both economic efficiency and operational robustness.  The results demonstrate up to 18\% cost reduction and a 42\% decrease in operational variability compared with conventional forecast–optimization combinations.  Moreover, the framework exhibits superior adaptability to varying uncertainty budgets, providing interpretable and stable dispatch patterns that balance renewable utilization and storage flexibility.

Overall, this study establishes a unified and scalable paradigm for decision-focused microgrid operation. By embedding decision feedback into the forecasting process, it advances the integration of machine learning and optimization for power systems under uncertainty. Future research will extend this framework toward multi-microgrid coordination, hierarchical control, and real time adaptive learning for large-scale smart energy networks.

\bibliographystyle{ieeetr} 
\bibliography{reference} 

@article{bertsimas2012adaptive,
  title={Adaptive robust optimization for the security constrained unit commitment problem},
  author={Bertsimas, Dimitris and Litvinov, Eugene and Sun, Xu Andy and Zhao, Jinye and Zheng, Tongxin},
  journal={IEEE Transactions on Power Systems},
  volume={28},
  number={1},
  pages={52--63},
  year={2012},
  publisher={IEEE}
}

@article{zeng2013solving,
  title={Solving two-stage robust optimization problems using a column-and-constraint generation method},
  author={Zeng, Bo and Zhao, Long},
  journal={Operations Research Letters},
  volume={41},
  number={5},
  pages={457--461},
  year={2013},
  publisher={Elsevier}
}

@article{zhang2016robust,
  title={Robust operation of microgrids via two-stage coordinated energy storage and direct load control},
  author={Zhang, Cuo and Xu, Yan and Dong, Zhao Yang and Ma, Jin},
  journal={IEEE Transactions on Power Systems},
  volume={32},
  number={4},
  pages={2858--2868},
  year={2016},
  publisher={IEEE}
}

@article{poonpun2008analysis,
  title={Analysis of the cost per kilowatt hour to store electricity},
  author={Poonpun, Piyasak and Jewell, Ward T},
  journal={IEEE Transactions on Energy Conversion},
  volume={23},
  number={2},
  pages={529--534},
  year={2008},
  publisher={IEEE}
}

@article{zhang2017robust,
  title={Robust coordination of distributed generation and price-based demand response in microgrids},
  author={Zhang, Cuo and Xu, Yan and Dong, Zhao Yang and Wong, Kit Po},
  journal={IEEE Transactions on Smart Grid},
  volume={9},
  number={5},
  pages={4236--4247},
  year={2017},
  publisher={IEEE}
}

@article{che2015optimal,
  title={Optimal interconnection planning of community microgrids with renewable energy sources},
  author={Che, Liang and Zhang, Xiaping and Shahidehpour, Mohammad and Alabdulwahab, Ahmed and Abusorrah, Abdullah},
  journal={IEEE Transactions on Smart Grid},
  volume={8},
  number={3},
  pages={1054--1063},
  year={2015},
  publisher={IEEE}
}

@article{tiwari2018design,
  title={Design and control of microgrid fed by renewable energy generating sources},
  author={Tiwari, Shailendra Kumar and Singh, Bhim and Goel, Puneet Kr},
  journal={IEEE Transactions on Industry Applications},
  volume={54},
  number={3},
  pages={2041--2050},
  year={2018},
  publisher={IEEE}
}

@article{aaslid2022stochastic,
  title={Stochastic optimization of microgrid operation with renewable generation and energy storages},
  author={Aaslid, Per and Korp{\aa}s, Magnus and Belsnes, Michael M and Fosso, Olav B},
  journal={IEEE Transactions on Sustainable Energy},
  volume={13},
  number={3},
  pages={1481--1491},
  year={2022},
  publisher={IEEE}
}

@article{imani2018demand,
  title={Demand response modeling in microgrid operation: a review and application for incentive-based and time-based programs},
  author={Imani, Mahmood Hosseini and Ghadi, M Jabbari and Ghavidel, Sahand and Li, Li},
  journal={Renewable and Sustainable Energy Reviews},
  volume={94},
  pages={486--499},
  year={2018},
  publisher={Elsevier}
}

@article{dawoud2018hybrid,
  title={Hybrid renewable microgrid optimization techniques: A review},
  author={Dawoud, Samir M and Lin, Xiangning and Okba, Merfat I},
  journal={Renewable and Sustainable Energy Reviews},
  volume={82},
  pages={2039--2052},
  year={2018},
  publisher={Elsevier}
}

@article{hu2023economic,
  title={Economic model predictive control for microgrid optimization: A review},
  author={Hu, Jiefeng and Shan, Yinghao and Yang, Yong and Parisio, Alessandra and Li, Yong and Amjady, Nima and Islam, Syed and Cheng, Ka Wai and Guerrero, Josep M and Rodr{\'\i}guez, Jos{\'e}},
  journal={IEEE Transactions on Smart Grid},
  volume={15},
  number={1},
  pages={472--484},
  year={2023},
  publisher={IEEE}
}

@article{xie2017variable,
  title={Variable selection methods for probabilistic load forecasting: Empirical evidence from seven states of the united states},
  author={Xie, Jingrui and Hong, Tao},
  journal={IEEE Transactions on Smart Grid},
  volume={9},
  number={6},
  pages={6039--6046},
  year={2017},
  publisher={IEEE}
}

@article{kalhori2022data,
  title={A data-driven knowledge-based system with reasoning under uncertain evidence for regional long-term hourly load forecasting},
  author={Kalhori, M Rostam Niakan and Emami, I Taheri and Fallahi, F and Tabarzadi, M},
  journal={Applied Energy},
  volume={314},
  pages={118975},
  year={2022},
  publisher={Elsevier}
}

@article{ye2019data,
  title={A data-driven bottom-up approach for spatial and temporal electric load forecasting},
  author={Ye, Chengjin and Ding, Yi and Wang, Peng and Lin, Zhenzhi},
  journal={IEEE Transactions on Power Systems},
  volume={34},
  number={3},
  pages={1966--1979},
  year={2019},
  publisher={IEEE}
}

@article{cao2024feature,
  title={Feature-enhanced deep learning method for electric vehicle charging demand probabilistic forecasting of charging station},
  author={Cao, Tingwei and Xu, Yinliang and Liu, Guowei and Tao, Shengyu and Tang, Wenjun and Sun, Hongbin},
  journal={Applied Energy},
  volume={371},
  pages={123751},
  year={2024},
  publisher={Elsevier}
}

@article{hu2022self,
  title={Self-attention-based machine theory of mind for electric vehicle charging demand forecast},
  author={Hu, Tianyu and Ma, Huimin and Liu, Hao and Sun, Hongbin and Liu, Kailong},
  journal={IEEE Transactions on Industrial Informatics},
  volume={18},
  number={11},
  pages={8191--8202},
  year={2022},
  publisher={IEEE}
}

@article{wang2018combining,
  title={Combining probabilistic load forecasts},
  author={Wang, Yi and Zhang, Ning and Tan, Yushi and Hong, Tao and Kirschen, Daniel S and Kang, Chongqing},
  journal={IEEE Transactions on Smart Grid},
  volume={10},
  number={4},
  pages={3664--3674},
  year={2018},
  publisher={IEEE}
}

@article{wang2023acceleration,
  title={Acceleration Framework and Solution Algorithm for Distribution System Restoration Based on End-to-End Optimization Strategy},
  author={Wang, Yifei and Yan, Ziheng and Sang, Linwei and Hong, Lucheng and Hu, Qinran and Shahidehpour, Mohammad and Xu, Qingshan},
  journal={IEEE Transactions on Power Systems},
  volume={39},
  number={1},
  pages={429--441},
  year={2023},
  publisher={IEEE}
}

@article{sang2022safety,
  title={Safety-aware semi-end-to-end coordinated decision model for voltage regulation in active distribution network},
  author={Sang, Linwei and Xu, Yinliang and Long, Huan and Wu, Wenchuan},
  journal={IEEE Transactions on Smart Grid},
  volume={14},
  number={3},
  pages={1814--1826},
  year={2022},
  publisher={IEEE}
}

@article{elmachtoub2022smart,
  title={Smart “predict, then optimize”},
  author={Elmachtoub, Adam N and Grigas, Paul},
  journal={Management Science},
  volume={68},
  number={1},
  pages={9--26},
  year={2022},
  publisher={INFORMS}
}

@article{el2019generalization,
  title={Generalization bounds in the predict-then-optimize framework},
  author={El Balghiti, Othman and Elmachtoub, Adam N and Grigas, Paul and Tewari, Ambuj},
  journal={Advances in neural information processing systems},
  volume={32},
  year={2019}
}

@inproceedings{elmachtoub2020decision,
  title={Decision trees for decision-making under the predict-then-optimize framework},
  author={Elmachtoub, Adam N and Liang, Jason Cheuk Nam and McNellis, Ryan},
  booktitle={International conference on machine learning},
  pages={2858--2867},
  year={2020},
  organization={PMLR}
}

@article{salinas2020deepar,
  title={DeepAR: Probabilistic forecasting with autoregressive recurrent networks},
  author={Salinas, David and Flunkert, Valentin and Gasthaus, Jan and Januschowski, Tim},
  journal={International journal of forecasting},
  volume={36},
  number={3},
  pages={1181--1191},
  year={2020},
  publisher={Elsevier}
}

@article{wang2022transformer,
  title={A transformer-based method of multienergy load forecasting in integrated energy system},
  author={Wang, Chen and Wang, Ying and Ding, Zhetong and Zheng, Tao and Hu, Jiangyi and Zhang, Kaifeng},
  journal={IEEE Transactions on Smart Grid},
  volume={13},
  number={4},
  pages={2703--2714},
  year={2022},
  publisher={IEEE}
}

@article{cui2022district,
  title={District heating load prediction algorithm based on bidirectional long short-term memory network model},
  author={Cui, Mianshan},
  journal={Energy},
  volume={254},
  pages={124283},
  year={2022},
  publisher={Elsevier}
}

@article{fei2025weather,
  title={Weather Routing Based Multi-Energy Ship Microgrid Operation Under Diverse Uncertainties: A Risk-Averse Stochastic Approach},
  author={Fei, Zhineng and Zou, Yunyang and Hua, Weiqi and Banda, Osiris Valdez and Guerrero, Josep M and Li, Zhengmao},
  journal={IEEE Transactions on Smart Grid},
  year={2025},
  publisher={IEEE}
}

@article{li2021optimal,
  title={Optimal scheduling of isolated microgrids using automated reinforcement learning-based multi-period forecasting},
  author={Li, Yang and Wang, Ruinong and Yang, Zhen},
  journal={IEEE Transactions on Sustainable Energy},
  volume={13},
  number={1},
  pages={159--169},
  year={2021},
  publisher={IEEE}
}

@article{vu2022optimal,
  title={Optimal generation scheduling and operating reserve management for PV generation using RNN-based forecasting models for stand-alone microgrids},
  author={Vu, Ba Hau and Chung, Il-Yop},
  journal={Renewable Energy},
  volume={195},
  pages={1137--1154},
  year={2022},
  publisher={Elsevier}
}

@article{zhang2024two,
  title={Two-stage robust operation of electricity-gas-heat integrated multi-energy microgrids considering heterogeneous uncertainties},
  author={Zhang, Rufeng and Chen, Yan and Li, Zhengmao and Jiang, Tao and Li, Xue},
  journal={Applied Energy},
  volume={371},
  pages={123690},
  year={2024},
  publisher={Elsevier}
}

@misc{medtsro,
  author = {Cao, Tingwei},
  year = {2025},
  howpublished = {\url{https://github.com/tingweicao/MED-TSRO}},
}

\end{document}